# The Foucault pendulum as an example of motion on a pseudo-surface

D. H. Delphenich ([1])
Spring Valley, OH, USA

**Abstract:** The Foucault pendulum is shown to be an example of motion on a pseudo-surface, and the consequences of that are explored. In particular, its first and second fundamental forms are obtained, as well as its Gaussian and mean curvatures and the equations of its geodesics. However, a physical consideration that relates to the extension from space to space-time introduces a complication into the discussion of the geometry of the Foucault pseudo-surface that relates to the possible signatures of the space-time metric.

**1. Introduction.** – To some extent, the present study is an addendum to a more extensive discussion by the author [**1**] on the geometry of pseudo-hypersurfaces and how it relates to motion with a non-holonomic linear constraint that is defined by a single Pfaff equation. However, a brief review of the basic ideas and results of that previous article will be provided in order to make the present article more self-contained. Nonetheless, some prior exposure to the larger work would be helpful.

The concept of a pseudo-surface goes back to an 1889 paper [**2**] by the French abbot Pierre Adolphe Issaly, who was basically trying to apply the methods of the Pfaff equation to the geometry of surfaces. The main difference between a surface and a pseudo-surface relates to the degree of integrability of that Pfaff equation, namely, the maximum dimension of its integral submanifolds. Thus, a surface is defined by a completely-integrable Pfaff equation that has two-dimensional integral submanifolds, while a pseudo-surface has only integral curves.

Some time later (1899), Élie Cartan [**3**] introduced a method for obtaining the degree of integrability of a Pfaff equation that involved looking at a sequence of differential forms of increasing degree that are defined by exterior products of the 1-form (i.e., the Pfaffian) that defines the Pfaff equation and its exterior derivative. Ultimately, one gets a "normal" form for the Pfaffian that generalizes what one would get in the simplest case where it is exact.

The geometry of a pseudo-surface is closely analogous to the geometry of a surface and takes advantage of the fact that as long as one is defining things in terms of tangent and cotangent objects, it does not necessarily matter if they are tangent and cotangent to an integral surface. In particular, one can still define the first and second fundamental forms, and thus the principal curvatures, the Gaussian and mean curvatures, and the geodesic equations.

Hence, the basic structure of this article will be to first define a pseudo-surface by a Pfaff equation and discuss the integrability of the Pfaff equation in the two-dimensional case, and then in section **3**, we will review the relevant constructions in the theory of the geometry of surfaces ([2]). Particular emphasis will be given to the role of frame fields that are adapted to the surface as a way of getting the geometric objects that it defines. In section **4**, the geometric constructions that were introduced in the context of surfaces will then be defined in the case of pseudo-surfaces using the method of adapted frame fields. The next section reviews the physical basis for the Foucault pendulum and presents its equations of motion as an example of motion with a non-holonomic constraint. The treatment of that problem is closely based in the treatment that Georg Hamel gave it in [**6**]. Finally, the relationship between the mathematical modelling of the Foucault pendulum and the geometry of pseudo-surfaces is discussed, and we conclude with a brief summary of the results.

---

([1]) E-Mail: feedback@neo-classical-physics.info. Website: neo-classical-physics.info
([2]) A good classical reference on the geometry of curves and surfaces is Eisenhart [**4**], while a more modern one that consistently uses Cartan's method of moving frames and differential forms is O'Neill [**5**].

As we will see, the physics of the problem actually introduces a complication into the discussion of the geometry of the pseudo-surface that is defined by the Foucault pendulum. It relates to the extension of the metric on "space" to a metric on "space-time" and the fact that naively extending the Euclidian metric on space to a Euclidian metric on space-time would have no physical basis, even though it would make the geometry of the pseudo-surface more straightforward. A so-called "Galilean extension" of the Euclidian metric to a degenerate Euclidian metric would render the first fundamental form of the pseudo-surface non-invertible, and thus useless for "raising an index" on the second fundamental form in order to obtain its eigenvalues. On the other hand, a Minkowskian extension, in the spirit of special relativity, would make the first fundamental form non-degenerate, so one could still get the eigenvalues of the second fundamental form, but they would be imaginary.

**2. Definition of a pseudo-surface.** – A surface in $\mathbb{R}^3$ can be represented by either an embedding of a two-dimensional submanifold (i.e., as a *locus*) or as a level surface of a function on $\mathbb{R}^3$ (i.e., as an *envelope*). Tangent planes to the surface will be defined in either case as long as the maps in question are differentiable. In the case of an embedding, the tangent planes to the surface are the images of the tangent planes to its parameter space under the differential of the embedding map. In the case of the level surfaces of a differentiable function $f$ on $\mathbb{R}^3$, one can define a plane at every point where the function is defined, namely, the plane that is annihilated by *df*. Hence, the distribution of planes will admit more than one surface that is tangent to them, namely, $\mathbb{R}^3$ will be "foliated" by the level surfaces of *f* like the pages of the book. One should be cautioned, though, that there might be singular points where *df* vanishes.

By contrast, if one defines only a field of planes on $\mathbb{R}^3$ (i.e., a differential system of codimension one) then the issue of whether there exists an embedding of a two-dimensional manifold that will be tangent to some of those planes at its image points is a question of the integrability of the differential system. A *pseudo-surface* is simply a non-integrable differential system of codimension one on $\mathbb{R}^3$.

One way of defining the field of planes on $\mathbb{R}^3$ is to define a non-zero 1-form $\theta$ that generalizes *df*, which is also called a *Pfaffian* form, after the German mathematician Johann Friedrich Pfaff (1765-1825) who published a landmark treatise on the integration of first-order partial differential equations that opened up an entire branch of analysis and geometry that produced some of the most useful advances in those fields throughout the nineteenth and twentieth centuries. The exterior differential equation $\theta = 0$ is then referred to as the *Pfaff equation.*

First of all, the Pfaff equation defines a plane at every point $x \in \mathbb{R}^3$ where $\theta$ is defined, namely, the *annihilating plane* of the 1-form. That is, it is the set of all tangent vectors $\mathbf{v} \in T_x \mathbb{R}^3$ such that $\theta(\mathbf{v}) = 0$. In the case where the Pfaffian $\theta = df$ for some differentiable function $f$ on $\mathbb{R}^3$, the annihilating planes will be tangent planes to the level surfaces of $f$, which are the sets of all $x$ that go to the same value under $f(x)$.

More generally, the 1-form $\theta$ does not have to be *exact* (i.e., take the form $df$ for some $f$), or even *closed* (i.e., its exterior derivative $d_\wedge \theta = 0$), but can take one of three *normal forms*, namely:

$$\theta = d\mu, \quad \lambda\, d\mu, \quad d\phi + \lambda\, d\mu, \tag{2.1}$$

respectively, in which $\lambda$, $\mu$, and $\phi$ are differentiable functions (i.e., 0-forms). (The fact that there are three normal forms is specific to the three dimensions of the manifold on which $\theta$ is defined.)

Upon taking exterior derivatives, one sees that the three normal forms give:

$$d_\wedge \theta = 0, \quad d\lambda \wedge d\mu, \quad d\lambda \wedge d\mu, \tag{2.2}$$

respectively. Furthermore, when one forms the 3-form $\theta \wedge d_\wedge \theta$, which we shall refer to as the *Frobenius 3-form*, after the German mathematician Ferdinand Georg Frobenius (1849-1917), who continued the work of Pfaff and proved a powerful theorem for the complete integrability of the Pfaff equation, one will get:

$$\theta \wedge d_\wedge \theta = 0, \quad 0, \quad d\phi \wedge d\lambda \wedge d\mu, \tag{2.3}$$

respectively.

An *integral submanifold* of the Pfaff equation is a differentiable embedding $x : S \to \mathbb{R}^3$ of a one or two-dimensional manifold $S$ in $\mathbb{R}^3$ such that the images of the tangent lines or planes to $S$ under the differential map $dx$ are linear subspaces of the annihilating planes of $\theta$; hence, an integral submanifold can be a differentiable curve or surface in $\mathbb{R}^3$ depending upon whether a tangent line or tangent plane is annihilated by $\theta$. The *degree of integrability* of the Pfaff equation is the maximum dimension of the integral submanifolds that exist, so in the present case that will be one or two. One-dimensional integral submanifolds always exist, but the issue of the *complete integrability* of the Pfaff equation (i.e., the existence of integral submanifolds of codimension one, which is to say dimension two, in this case) is the subject of *Frobenius's theorem*. In one form, it says that $\theta = 0$ is completely integrable iff $\theta \wedge d_\wedge \theta$ vanishes, which will happen when $\theta$ takes the normal forms $d\mu$ and $\lambda\, d\mu$, but not when it takes the normal form $d\phi + \lambda\, d\mu$. In the case of $\lambda\, d\mu$, one often says that $\theta$ admits an "integrating factor," and in either of the completely-integrable cases, the integral submanifolds will be level surfaces of the function $\mu$. ($\lambda$ is assumed to be everywhere non-zero.)

It is the non-completely-integrable case that defines a pseudo-surface. However, as was just mentioned, integral curves will exist; that is, differentiable curves $x(s)$ in $\mathbb{R}^3$ whose tangent vectors $\mathbf{t}(s)$ all lie in the annihilating planes of $\theta$, which can be expressed in the form:

$$\theta(\mathbf{t}(s)) = 0, \quad \text{for all } s. \tag{2.4}$$

If $\mathbb{R}^3$ is given the canonical coordinate system $(x^1, x^2, x^3)$ and the natural frame field $\{\partial_1, \partial_2, \partial_3\}$ that is defined by the directional derivative operators $\partial_i = \partial / \partial x_i$ along the coordinate lines, as well as its reciprocal natural coframe field $\{dx^1, dx^2, dx^3\}$, then the curve $x(s)$, its tangent vectors $\mathbf{t}(s)$, and the Pfaffian form $\theta$ can be represented in the form of systems of equations ($i = 1, 2, 3$, summation convention in effect):

$$x^i = x^i(s), \qquad \mathbf{t}(s) = \frac{dx^i}{ds}(s)\partial_i, \qquad \theta = \theta_i(x)\, dx^i. \tag{2.5}$$

The condition (2.4) can then be represented in the form:

$$\theta_i(x^j(s))\frac{dx^i}{ds}(s) = 0. \tag{2.6}$$

In the completely-integrable case when $\theta = df$, that will reduce to:

$$\frac{\partial f}{\partial x^i}\frac{dx^i}{ds} = \frac{df}{ds} = 0, \tag{2.7}$$

which can also be expressed by saying that the directional derivative $\mathbf{t} f = df(\mathbf{t})$ vanishes for all $s$, so $f$ will be constant along the curve, which is consistent with the idea that the curve lies in a level surface of $f$.

Before we focus on the geometry of pseudo-surfaces, we shall first recall some constructions in the geometry of surfaces that can be generalized directly. The key to that generalization is in the replacement of a holonomic adapted frame field with an anholonomic one, i.e., the replacement of an adapted coordinate system with an adapted frame field that cannot be a natural one.

**3. Geometry of surfaces.** – We shall recall some of the elementary notions from the geometry of surfaces and express them in a form that can be directly adapted to the demands of pseudo-surfaces.

*a. Frame fields adapted to surfaces.* – If a surface in $\mathbb{R}^3$ is defined by an embedding $x : S \to \mathbb{R}^3$, $(u^a) \mapsto x^i(u^a)$ then one can define a (local) frame field on $S$ by way of the "push-forward" of the natural frame field $\{\partial_a = \partial / \partial u^a\}$ over a coordinate chart $(U, u^a)$ on $S$:

$$\mathbf{e}_a(u) = x_* \delta_a = x^i_{,i}(u) \partial_i, \qquad x^i_{,j}(u) = \frac{\partial x^i}{\partial u^a}(u). \tag{3.1}$$

Hence, the effect of $\mathbf{e}_a(u)$ on a differentiable function $f(x)$ as a directional derivative is the same as $\partial / \partial u^a$:

$$\mathbf{e}_a(u) f = \frac{\partial x^i}{\partial u^a} \frac{\partial f}{\partial x^i} = \frac{\partial f}{\partial u^a}, \tag{3.2}$$

when $x = x(u)$.

If $\mathbf{N} = N^i \partial_i$ is a normal vector field to $S$ then we can complete the frame field $\mathbf{e}_a(u)$ on $S$, which spans its tangent spaces, to a 3-frame field:

$$\mathbf{e}_i(u) = x^j_i(u) \partial_j, \tag{3.3}$$

in which we have defined the invertible matrix:

$$x^j_i(u) = \left[ \frac{\partial x^j}{\partial u^a}(u) \ \vdots \ N^j(x(u)) \right]. \tag{3.4}$$

For the sake of specificity, we give $\mathbb{R}^3$ the Euclidian metric and define $\mathbf{N}$ to be the unit vector field that is orthogonal to both $\mathbf{e}_1$ and $\mathbf{e}_2$ and makes the 3-frame $\{\mathbf{e}_1, \mathbf{e}_2, \mathbf{N}\}$ right-handed. We will make the frame $\mathbf{e}_a(u)$ orthonormal later when we define a metric on $S$. Hence, the 3-frame field $\mathbf{e}_i(u)$ is adapted to $S$ in the sense that $\mathbf{e}_a(u)$ span the tangent spaces to its image in $\mathbb{R}^3$, while $\mathbf{e}_3(u)$ is normal to those tangent spaces.

The coframe field $\theta^i$ that is reciprocal to $\mathbf{e}_i(u)$ is defined to make $\theta^i(\mathbf{e}_j) = \delta^i_j$, so it can be represented in terms of the natural coframe $dx^i$ on $\mathbb{R}^3$ as:

$$\theta^i = \tilde{x}^i_j dx^j, \tag{3.5}$$

which gives the 1-form:

$$N \equiv \theta^3 = \tilde{x}^3_j dx^j \tag{3.6}$$

the property that:

$$N(\mathbf{N}) = \tilde{x}^3_j x^j_3 = 1. \tag{3.7}$$

Thus, we can regard $N$ as the unit normal 1-form to $S$. If $S$ is defined by a level surface of a differentiable function $f$ on $\mathbb{R}^3$ then we can define $N$ to be the normalization of $df$:

$$N = \frac{1}{\|df\|} df = \frac{1}{\|df\|} \frac{\partial f}{\partial x^i} dx^i, \qquad \text{so} \qquad N_i = \frac{1}{\|df\|} \frac{\partial f}{\partial x^i}. \tag{3.8}$$

Since the metric-dual of *df* is the gradient of *f*, one can then consistently regard **N** as the normalized gradient of *f*:

$$\mathbf{N} = \frac{1}{\|df\|}\nabla f = \left(\frac{1}{\|df\|}\delta^{ij}\frac{\partial f}{\partial x^j}\right)\frac{\partial}{\partial x^i}, \qquad \text{so} \qquad N^i = \frac{1}{\|df\|}\delta^{ij}\frac{\partial f}{\partial x^j}. \qquad (3.9)$$

If we differentiate the frame field $\mathbf{e}_i(u)$ then we will get:

$$d\mathbf{e}_i = dx_i^j \otimes \partial_j = \omega_i^j \otimes \mathbf{e}_j, \qquad (3.10)$$

in which we have introduced the matrix of 1-forms:

$$\omega_i^j = dx_i^k \, \tilde{x}_k^j, \qquad (3.11)$$

in which the tilde over a matrix will signify the inverse of the matrix from now on. Since $dx_i^k \, \tilde{x}_k^j = (\partial_k x_i^l \, \tilde{x}_l^j)dx^k$, we can also express $\omega_i^j$ in terms of the adapted coframe field in the form:

$$\omega_j^i = \omega_{jk}^i \, \theta^k, \qquad \text{with} \qquad \omega_{jk}^i = x_k^m \, \partial_m x_j^l \, \tilde{x}_l^i = (\mathbf{e}_k x_j^l)\tilde{x}_l^i. \qquad (3.12)$$

Since $\mathbf{e}_a$ has the effect of $\partial/\partial u^a$ and $x_a^i = \partial x^i / \partial u^a$, we will then have:

$$\omega_{ab}^i = \frac{\partial^2 x^k}{\partial u^a \, \partial u^b}\tilde{x}_k^i, \qquad (3.13)$$

which is automatically symmetric in the indices *ab* as long as the embedding *x* is twice continuously differentiable.

When the matrix $\omega_i^j$ is given, the equations (3.10) can be regarded as a system of three linear, first-order partial differential equations for the frame field $\mathbf{e}_i$ on *S* and will thus be regarded as the differential equations of that frame field.

The coframe field $\theta^i$ on *S* that is reciprocal to $\mathbf{e}_i$ will then have the differential equation:

$$d\theta^i = -\omega_j^i \otimes \theta^j, \qquad (3.14)$$

which follows from differentiating the defining equation (3.5) for $\theta^i$ and noting that since $x_k^i \, \tilde{x}_j^k = \delta_j^i$, differentiating that will give:

$$\omega_j^i = dx_k^i \, \tilde{x}_j^k = -x_k^i \, d\tilde{x}_j^k. \qquad (3.15)$$

One can restrict the 3-frame $\mathbf{e}_i$ to a 2-frame $\{\mathbf{e}_a, a = 1, 2\}$, which will span the tangent spaces to *S*, and with $\mathbf{e}_3 = \mathbf{N}$, the equations of the frame field can then be put into the form:

$$de_a = \omega_a^b \otimes e_b + \omega_a^3 \otimes N, \qquad de_3 = \omega_3^b \otimes e_b + \omega_3^3 \otimes N. \tag{3.16}$$

If the frame field $e_i$ is defined to be orthonormal then the matrix $\omega_j^i$ will be anti-symmetric, and the component $\omega_3^3$ will have to vanish.

Dually, one has:

$$d\theta^a = -\omega_b^a \otimes \theta^b - \omega_3^a \otimes N, \qquad dN = -\omega_b^3 \otimes \theta^b - \omega_3^3 \otimes N, \tag{3.17}$$

since $\theta^3 = N$.

One verifies that the symmetry of $x^i_{,a,b}$ and $f_{,a,b}$ in their lower indices implies that:

$$[e_a, e_b] = (x^j_{,a} \partial_j x^i_{,b} - x^j_{,b} \partial_j x^i_{,a}) \partial_i = (\partial_a x^i_{,b} - \partial_b x^i_{,a}) \partial_i = 0 \tag{3.18}$$

and

$$d_\wedge N = d_\wedge df = 0. \tag{3.19}$$

Thus, the adapted frame field $e_a$ on $S$ is *holonomic*, by definition.

The matrix of 1-forms $\omega_j^i$ can be regarded as the 1-form of the linear connection that makes the frame field $e_i$ parallel, and is often referred to as the *teleparallelism connection* then. The author has written some articles on the application of teleparallelism [**7**] to mechanics, as well as compiling an anthology [**8**] of his translations of early papers on its possible role in general relativity and the Einstein-Maxwell unification program.

*b. The fundamental forms of a surface.* – In the case of a surface $S$ (i.e., a level set of some $f$), one can define two fundamental forms (viz., symmetric, second-rank, covariant tensor fields) on $S$. The *first fundamental form* is simply the pull-back of the Euclidian metric $\delta$ on $\mathbb{R}^3$ by the embedding $x$:

$$g = x^* \delta, \tag{3.20}$$

so if $v$, $w$ are tangent vectors to $u \in S$ then:

$$g(v, w) = \delta(dx|_u(v), dx|_u(w)). \tag{3.21}$$

If $\{x^1, x^2, x^3\}$ is a coordinate system on $\mathbb{R}^3$ then $\delta$ will take the component form:

$$\delta = \delta_{ij} \, dx^i \, dx^j, \tag{3.22}$$

in which the multiplication of the basic 1-forms $dx^i$ and $dx^j$ is the symmetrized tensor product. If $\{u^1, u^2\}$ is a local coordinate system on $S$ then the component representation of $g$ will then take the form ($a, b = 1, 2$):

$$g = g_{ab} \, du^a \, du^b, \qquad g_{ab} = \delta_{ij} \frac{\partial x^i}{\partial u^a} \frac{\partial x^j}{\partial u^b}. \tag{3.23}$$

The second fundamental form defines the "shape" of $S$ in terms of the second differential of the function $f$ that defines $S$ as a level surface, although one must first normalize $df$ to a unit 1-form:

$$N = \frac{1}{\|df\|} df = N_i \, dx^i \qquad (N_i = \frac{1}{\|df\|} f_{,i}) . \tag{3.24}$$

One then differentiates the 1-form $N$ to obtain a second-rank tensor field:

$$dN = \frac{1}{\|df\|} \left[ d^2 f - \frac{1}{\|df\|^2} < d^2 f, df > \otimes df \right] = N_{i,j} \, dx^i \otimes dx^j , \tag{3.25}$$

in which:

$$N_{i,j} = \frac{1}{\|df\|} \left[ f_{,i,j} - \frac{1}{\|df\|^2} \delta^{kl} f_{,i,k} \, f_{,l} \, f_{,j} \right]. \tag{3.26}$$

In order to get the second fundamental form $H$ on $S$ from this, one pulls $N$ back using $dx$ ([1]):

$$H = - x^* \, dN = - \frac{\partial x^i}{\partial u^a} \frac{\partial x^j}{\partial u^b} N_{i,j} \, du^a du^b . \tag{3.27}$$

Hence, the local components of $H$ with respect to the natural coframe $du^a$ on the subset of $S$ over which it is defined will be:

$$H_{ab} = - \frac{\partial x^i}{\partial u^a} \frac{\partial x^j}{\partial u^b} N_{i,j} = - \frac{1}{\|df\|} \frac{\partial x^i}{\partial u^a} \frac{\partial x^j}{\partial u^b} \frac{\partial^2 f}{\partial x^i \partial x^j} . \tag{3.28}$$

The second term in (3.26) has dropped out because:

$$\frac{\partial x^j}{\partial u^b} \frac{\partial f}{\partial x^j} = \frac{\partial f}{\partial u^b} = 0 , \tag{3.29}$$

since the coordinates $u^b$ all describe points of a level surface of $f$. Note that $H_{ab}$ is symmetric in its indices, which follows from the fact that the second mixed partial derivatives of any twice-continuously differentiable function (such as $f$) will be symmetric in the derivatives.

One can also express the second fundamental form in terms of the second derivatives of $x$ by differentiating (3.29):

$$0 = \frac{\partial^2 x^j}{\partial u^a \partial u^b} \frac{\partial f}{\partial x^j} + \frac{\partial x^j}{\partial u^b} \frac{\partial^2 f}{\partial u^a \partial x^j} = \frac{\partial^2 x^i}{\partial u^a \partial u^b} \frac{\partial f}{\partial x^i} + \frac{\partial x^i}{\partial u^a} \frac{\partial x^j}{\partial u^b} \frac{\partial^2 f}{\partial u^a \partial u^b} = \frac{\partial^2 x^i}{\partial u^a \partial u^b} \frac{\partial f}{\partial x^i} - \|df\| H_{ab} ,$$

which makes:

---

([1]) This definition will imply the one in Eisenhart [**4**], as we shall see.

$$H_{ab} = \frac{1}{\|df\|} \frac{\partial^2 x^i}{\partial u^a \partial u^b} \frac{\partial f}{\partial x^i} = \frac{\partial^2 x^i}{\partial u^a \partial u^b} N_i . \qquad (3.30)$$

(This is the form that one finds in Eisenhart.)

Perhaps the reason for calling g and H the "fundamental" forms of a surface comes down to the fact that if one is given g and H, but not x and N, then as long as the tensor fields g and H satisfy certain integrability conditions (viz., the Gauss-Codazzi-Mainardi equations [**4, 5**]), one can solve for x and N, which will be unique, up to a rigid motion in space. Thus, as far as the problem of determining the "shape" of a surface is concerned, g and H contain a complete statement of its solution. Note that some surfaces (such as developable surfaces) can be deformed in many non-rigid ways without changing the metric, so in that sense, considering only the metric geometry does not define a surface, except in a somewhat coarse-grained way. That is because, according to Beez's theorem, it is only for Riemannian manifolds of dimension greater than two that isometries must be rigid motions. Thus, the study of flexible, inextensible curves and surfaces in space never leaves a single isometry class of such objects.

One can make a linear operator on $\mathbb{R}^2$ out of these components by raising an index using the metric g on S :

$$H^a_b = g^{ac} H_{cb} . \qquad (3.31)$$

Although $g^{ab}$ and $H_{ab}$ are both symmetric in their indices, $H^a_b$ does not have to be. One sees that most easily by using matrix notation for $g^{ab}$ and $H_{ab}$ and transposing their product:

$$[g\,H]^T = H^T g^T = H\,g , \qquad (3.32)$$

which does not have to equal $g\,H$, in general. However, that will be true when the frame field $\mathbf{e}_a$ is orthonormal since one will have $g^{ab} = \delta^{ab}$ in that case. More generally, it will be true iff $g^{ab} = \Omega^2 \delta^{ab}$ for some non-zero function $\Omega$; i.e., for all metrics that are conformally Euclidian.

If $H^a_b$ is still symmetric then it will have two (not-necessarily distinct or non-zero) real eigenvalues $\kappa_1$, $\kappa_2$, which are the *principal curvatures* of S, and the curves in S whose tangent vectors are the corresponding eigenvectors are the *lines of curvature.*

A *principal frame* for H is an orthonormal frame whose vectors are eigenvectors of $H^a_b$. In such a frame, the matrix $H^a_b$ (or $H_{ab}$, in the Euclidian case) will then take the form:

$$H^a_b = \begin{bmatrix} \kappa_1 & 0 \\ 0 & \kappa_2 \end{bmatrix} . \qquad (3.33)$$

If **v** is a vector in $\mathbb{R}^2$ and its components with respect to a principal frame are $(v^1, v^2)$ then one will have:

$$H(\mathbf{v}, \mathbf{v}) = H_{ab}\,v^a v^b = \kappa_1 (v^1)^2 + \kappa_2 (v^2)^2 , \qquad (3.34)$$

which can be regarded as the curvature of $S$ in the direction of $\mathbf{v}$ when $\mathbf{v}$ is a unit vector since that is true of a unit eigenvector. The equation $H(\mathbf{v}, \mathbf{v}) = 1$ then defines a quadric surface in each tangent space to each point of $S$ that is referred to as the *Dupin indicatrix*.

One can define the determinant and trace of the matrix $H_b^a$, which define the *Gaussian curvature* $K = \kappa_1 \kappa_2$ and *mean curvature* $\bar{\kappa} = \frac{1}{2}(\kappa_1 + \kappa_2)$ of $S$. Note that the vanishing of the Gaussian curvature requires only the vanishing of *at least one* of the principal curvatures, but not both of them. Thus, all developable surfaces have vanishing Gaussian curvature, even though they might have non-zero curvature in some direction. When the mean curvature vanishes, $S$ is called *minimal*.

In general, if $\mathbf{t}(s)$ is the unit tangent vector to a curve $u(s)$ in $S$ then:

$$\kappa(s) = H(\mathbf{t}(s)\,\mathbf{t}(s)) = H_{ab}\frac{du^a}{ds}\frac{du^b}{\partial s} = \frac{1}{\|df\|}\frac{\partial^2 f}{\partial x^i \partial x^j}\frac{dx^i}{ds}\frac{dx^j}{\partial s} \tag{3.35}$$

will be its curvature at each point, in the Frenet-Serret sense of the word "curvature," when $u(s)$ is a geodesic. In order to see that, we shall recall some of the basic ideas of the Frenet-Serret equations [**4, 5**] and specialize them to the case of curves on surfaces.

*c. The Frenet-Serret equations of a curve in space.* – The Frenet frame field along a twice-continuously-differentiable curve $x(s)$ in $E^3$ (viz. $\mathbb{R}^3$ with the Euclidian metric) that is parameterized by arc-length $s$ is defined by the following vector fields along that curve:

$$\mathbf{t}(s) = \left.\frac{dx}{ds}\right|_s, \quad \mathbf{n}(s) = \left\|\frac{d\mathbf{t}}{ds}\right\|^{-1}\left.\frac{d\mathbf{t}}{ds}\right|_s, \quad \mathbf{b}(s) = \mathbf{t} \times \mathbf{n}. \tag{3.36}$$

These unit vector fields are referred to as the *tangent* vector field $\mathbf{t}(s)$, the (principal) *normal* vector field $\mathbf{n}(s)$, and the *binormal* vector field $\mathbf{b}(s)$, respectively. They then define an orthonormal frame field along the curve $x(s)$.

The equations of motion for the Frenet frame field then become the system of first-order linear ordinary differential equations for the frame members that are called the *Frenet-Serret equations:*

$$\frac{d}{ds}\begin{bmatrix}\mathbf{t}\\\mathbf{n}\\\mathbf{b}\end{bmatrix} = \begin{bmatrix}0 & \kappa & 0\\-\kappa & 0 & -\tau\\0 & \tau & 0\end{bmatrix}\begin{bmatrix}\mathbf{t}\\\mathbf{n}\\\mathbf{b}\end{bmatrix}, \tag{3.37}$$

in which $\kappa(s)$ is the *curvature* of the curve $x(s)$, and $\tau(s)$ is its *torsion*. Those functions are then defined by:

$$\kappa = \left\|\frac{d\mathbf{t}}{ds}\right\|, \quad \tau = \left\|\frac{d\mathbf{b}}{ds}\right\|. \tag{3.38}$$

If one renames the orthonormal frame field $\mathbf{e}_1 = \mathbf{t}$, $\mathbf{e}_2 = \mathbf{n}$, $\mathbf{e}_3 = \mathbf{b}$, and defines the matrix:

$$\varpi_j^i = \begin{bmatrix} 0 & \kappa & 0 \\ -\kappa & 0 & -\tau \\ 0 & \tau & 0 \end{bmatrix} \quad (3.39)$$

then (3.37) can be expressed more concisely as:

$$\frac{d\mathbf{e}_i}{ds} = \varpi_i^j \mathbf{e}_j . \quad (3.40)$$

This is the general form for the differential equations of frame fields along curves in space, even when they are not orthonormal, in which case the matrix $\varpi_i^j$ will belong to a Lie algebra of matrices that is different from the Lie algebra $\mathfrak{so}(3)$ of infinitesimal rotations. Due to the close formal similarity of the equations (3.40) for the frame field along the curve and the equations (3.10) for a frame field on a surface, when we get to curves on surfaces, it will be important to have distinct notations for the two matrices $\varpi_i^j$ and $\omega_i^j$, respectively.

One sees that the curvature originates in the second derivative of $x$ ($s$) with respect to arc-length, while the torsion originates in the third derivative. That follows from the fact that:

$$\tau \mathbf{b} = \frac{d\mathbf{b}}{ds} = \frac{d}{ds}(\mathbf{t} \times \mathbf{n}) = \frac{d\mathbf{t}}{ds} \times \mathbf{n} + \mathbf{t} \times \frac{d\mathbf{n}}{ds} = \kappa \mathbf{n} \times \mathbf{n} + \mathbf{t} \times \frac{d}{ds}\left(\frac{1}{\kappa}\frac{d\mathbf{t}}{ds}\right) = \mathbf{t} \times \frac{d}{ds}\left(\frac{1}{\kappa}\frac{d^2 x}{ds^2}\right).$$

*d. Curves on surfaces.* – By definition, a curve $x$ ($s$) in $\mathbb{R}^3$ will lie on the surface $S$ iff its tangent vector $\mathbf{t}$ ($s$) always lies in a tangent plane to $S$. Hence, since each tangent plane is annihilated by the normal 1-form $df$, one must have:

$$df|_{x(s)}(\mathbf{t}(s)) = 0 \qquad \text{for every } s. \quad (3.41)$$

If $\{\mathbf{e}_1, \mathbf{e}_2\}$ is a frame field on $S$ that is adapted to $S$ then there will exist a unique pair of components $\upsilon^1(s)$, $\upsilon^2(s)$ for each $s$ that makes:

$$\mathbf{t}(s) = \upsilon^a(s) \mathbf{e}_a(s), \quad (3.42)$$

in which $\mathbf{e}_a$ ($s$) is defined by $\mathbf{e}_a$ ($x(s)$). Hence, if $\mathbf{t}$ ($s$) = $t^i(s) \partial_i$ with respect to the natural frame on $\mathbb{R}^3$ then:

$$t^i = x^i_{,a} \upsilon^a = \frac{\partial x^i}{\partial u^a} \frac{du^a}{ds}. \quad (3.43)$$

If we differentiate $\mathbf{t}$ ($s$) with respect to $s$ then we will get:

$$\frac{d\mathbf{t}}{ds} = \frac{d\upsilon^a}{ds} \mathbf{e}_a + \upsilon^a \frac{d\mathbf{e}_a}{ds}. \quad (3.44)$$

We already have equations for $de_a$, namely, (3.16). In order to get from $de_a$ to $de_a / ds$, we simply evaluate the 1-form $de_a$ on $\mathbf{t}$:

$$\frac{d\mathbf{e}_a}{ds} = \omega_a^j(\mathbf{t})\mathbf{e}_j = \omega_a^b(\mathbf{t})\mathbf{e}_b + \omega_a^3(\mathbf{t})\mathbf{N}. \tag{3.45}$$

This clearly has a tangential component and a normal one. Thus:

$$\kappa \mathbf{n} = \left[\frac{dv^a}{ds} + \omega_b^a(\mathbf{t})v^b\right]\mathbf{e}_a + \left[\omega_b^3(\mathbf{t})v^b\right]\mathbf{N}. \tag{3.46}$$

This has a tangential part and a normal part, as well:

$$\kappa \mathbf{n}_t = \kappa_g = \left[\frac{dv^a}{ds} + \omega_b^a(\mathbf{t})v^b\right]\mathbf{e}_a, \qquad \kappa_N = \kappa \mathbf{n}_\perp = [\omega_b^3(\mathbf{t})v^b]\mathbf{N}, \tag{3.47}$$

in which the vector field $\kappa_g$ ($s$) along $x$ ($s$) is referred to as the *geodesic curvature* of the curve, while $\kappa_N$ ($s$) is its *normal curvature*. The geodesic curvature is then the projection of $\kappa$ ($s$) $\mathbf{n}$ ($s$) onto the tangent space to $S$ at $x$ ($s$). One then has a direct sum decomposition $[\mathbf{N}]_x \oplus T_xS$ of each tangent space $T_x\mathbb{R}^3$ in $\mathbb{R}^3$ at each point of $S$, which is the orthogonal decomposition since $\mathbf{N}$ was defined to be orthogonal to $T$ ($S$), so one can express the vector field $\kappa \mathbf{n}$ in the form:

$$\kappa \mathbf{n} = \kappa_g + \kappa_N. \tag{3.48}$$

To see how the curvature of $x$ ($s$) relates to the second fundamental form, one starts with the fact that $\mathbf{t}$ ($s$) is assumed to be annihilated by $df$. Differentiating $df(\mathbf{t})$ with respect to $s$ will give:

$$0 = \frac{d}{ds}[df(\mathbf{t})] = \frac{d(df)}{ds}(\mathbf{t}) + df\left(\frac{d\mathbf{t}}{ds}\right) = d^2f(\mathbf{t},\mathbf{t}) + \kappa\, df(\mathbf{n}),$$

which gives:

$$\kappa\, df(\mathbf{n}) = -d^2f(\mathbf{t},\mathbf{t}). \tag{3.49}$$

If we divide both sides of this by $\|df\|$ then we will get:

$$\kappa N(\mathbf{n}) = H(\mathbf{t},\mathbf{t}). \tag{3.50}$$

The number $N(\mathbf{n})$ represents the cosine of the angle between the unit normal $\mathbf{n}$ to the curve and the unit normal $\mathbf{N}$ to the surface, which is the metric-dual to the covector field $N$:

$$\mathbf{N} = N^i \partial_i, \qquad N^i = \delta^{ij} N_j. \tag{3.51}$$

Since $\mathbf{n}_\perp = \langle \mathbf{N}, \mathbf{n} \rangle \mathbf{N}$, from (3.50) and the second equation in (3.47), we will have:

$$\omega_b^3(\mathbf{t})\upsilon^b = H(\mathbf{t}, \mathbf{t}),  \qquad (3.52)$$

which can be written:

$$\omega_{(bc)}^3 \upsilon^b \upsilon^c = H_{bc} \upsilon^b \upsilon^c, \qquad (3.53)$$

in which we have symmetrized the lower indices of $\omega_{bc}^3$, since $\upsilon^b \upsilon^c$ is symmetric in those indices:

$$\omega_{(bc)}^3 = \tfrac{1}{2}(\omega_{bc}^3 + \omega_{cb}^3). \qquad (3.54)$$

However, since the curve in $S$ that gave us $\mathbf{t}$ was arbitrary, we can then say that:

$$\omega_{(bc)}^3 = H_{bc}. \qquad (3.55)$$

Thus, the connection 1-form $\omega_j^i$ already contains the information required to define the second fundamental form of the surface. One can also see that more directly by recalling that $H$ is defined by starting from:

$$dN = d\theta^3 = -\omega_{bc}^3 \, \theta^b \otimes \theta^c \qquad (3.56)$$

and symmetrizing it.

*e. Geodesics on surfaces.* – The number $N(\mathbf{n})$ will be positive unity iff the two vectors $\mathbf{n}$ and $\mathbf{N}$ coincide, which is one way of defining a *geodesic* on the surface, i.e., a curve on $S$ whose unit normal coincides with the unit normal to the surface. In such a case, one will have that the curvature of the geodesic satisfies the equation:

$$\kappa = H(\mathbf{t},\mathbf{t}), \qquad (3.57)$$

which is what was asserted above.

Clearly, $x(s)$ will be a geodesic iff the geodesic curvature vanishes. That will then give rise to the system of ordinary differential equations for $x^i(s)$:

$$0 = \frac{d\upsilon^a}{ds} + \omega_{(bc)}^a \upsilon^b \upsilon^c, \qquad (3.58)$$

which are the geodesic equations on $S$. Since $\mathbf{e}_a$ is the push-forward of a natural frame field on $S$, we can write:

$$\upsilon^a(s) = \frac{du^a}{ds}, \qquad (3.59)$$

and the geodesic equations can also be regarded as differential equations for the curve $u^a(s)$ in $S$:

$$0 = \frac{d^2 u^a}{ds^2} + \omega_{(bc)}^a \frac{du^b}{ds} \frac{du^c}{ds}. \qquad (3.60)$$

*f. Motion on surfaces.* – So far, we have been implicitly parameterizing our curves on *S* by arc length in order to make **t** a unit vector. In order to address curves that are parameterized by time *t* instead, we must make a few minor adjustments. First of all, if the reparameterization takes the form $s = s(t)$, which is assumed to be invertible and differentiable, along with its inverse, then the velocity of the curve $x(s)$ will get rescaled by the *speed* of the curve:

$$v(t) = \frac{ds}{dt};  \qquad (3.61)$$

that is:

$$\mathbf{v}(t) = \frac{dx}{dt} = \frac{ds}{dt}\frac{dx}{ds} = v(t)\,\mathbf{t}(s(t)). \qquad (3.62)$$

Although this is merely a rescaling of a tangent vector that does not change its direction, the same thing cannot be said of the acceleration:

$$\mathbf{a}(t) = \frac{d\mathbf{v}}{dt} = \frac{d}{dt}(v\mathbf{t}) = \frac{dv}{dt}\mathbf{t} + v\frac{d\mathbf{t}}{dt} = \dot{v}\mathbf{t} + v^2\frac{d\mathbf{t}}{ds},$$

which gives:

$$\mathbf{a}(t) = \dot{v}\mathbf{t} + \kappa v^2 \mathbf{n}. \qquad (3.63)$$

That would be a rescaling of $\kappa\mathbf{n}$ iff the curve $x(t)$ had constant speed.

If one represents $\kappa$ as $1/r$, where *r* is the instantaneous radius of curvature of the curve, then one will see that (3.63) has the same form as the decomposition of acceleration for circular motion into a tangential and a centripetal component, except that the sign of the centripetal term $(v^2/r)\mathbf{n}$ has changed; i.e., the vector **n** must point *towards* the instantaneous center, not away from it, unlike the instantaneous unit radius vector. One can also say that if motion on a circle is motion that is *constrained* to a circle then the centripetal acceleration is the contribution to the total acceleration that comes from the constraint itself.

When the curve $x(s)$ lies on a surface *S*, from (3.63), the normal component of **a** will be:

$$N(\mathbf{a}) = \kappa v^2 N(\mathbf{n}) = v^2 H(\mathbf{t},\mathbf{t}),$$

so

$$N(\mathbf{a}) = H(\mathbf{v},\mathbf{v}) = \omega^3_{(ab)} v^a v^b, \qquad (3.64)$$

which is just a rescaling of (3.50) and (3.53), resp.

From (3.63), **a** will have a tangential component of:

$$\mathbf{a}_t = \dot{v}\mathbf{t} + v^2 \kappa_g. \qquad (3.65)$$

One has:

$$v^2 \kappa_g = \left[v^2\frac{dv^a}{ds} + v^2\omega^a_b(\mathbf{t})v^b\right]\mathbf{e}_a = \left[v\frac{d}{dt}\left(\frac{du^a}{ds}\right) + \omega^a_{(bc)}\frac{du^b}{dt}\frac{du^b}{dt}\right]\mathbf{e}_a$$

$$= \left[ v \frac{d}{dt}\left(\frac{1}{v}\frac{du^a}{dt}\right) + \omega^a_{(bc)} \frac{du^b}{dt}\frac{du^b}{dt} \right] \mathbf{e}_a = \left[ -\frac{\dot{v}}{v}\frac{du^a}{dt} + \frac{d^2 u^a}{dt^2} + \omega^a_{(bc)} \frac{du^b}{dt}\frac{du^b}{dt} \right] \mathbf{e}_a$$

$$= -\dot{v}\mathbf{t} + \left[ \frac{d^2 u^a}{dt^2} + \omega^a_{(bc)} \frac{du^b}{dt}\frac{du^b}{dt} \right] \mathbf{e}_a \ .$$

When one adds this to $\dot{v}\mathbf{t}$, one will get the total tangential component of the acceleration:

$$\mathbf{a}_t = \left[ \frac{d^2 u^a}{dt^2} + \omega^a_{(bc)} \frac{du^b}{dt}\frac{du^b}{dt} \right] \mathbf{e}_a , \qquad (3.66)$$

and the condition for $u^a(t)$ to be a geodesic, when it is parameterized by time, will be:

$$\frac{d^2 u^a}{dt^2} + \omega^a_{(bc)} \frac{du^b}{dt}\frac{du^b}{dt} = 0 , \qquad (3.67)$$

which is a reparameterization of (3.60).

**4. Geometry of pseudo-surfaces.** – Many of the constructions in the previous section can be carried over to pseudo-surfaces by the simple replacement $\partial x^i / \partial u^a$ with $x^i_a$ and $df$ with a Pfaffian form $N$. The only things that will no longer exist are the embedding $x$ and the function $f$.

*a. Frame fields adapted to pseudo-surfaces.* – This time, we shall start with the unit 1-form $N$ as our way of defining tangent planes. That is, any vector $\mathbf{v}$ in the plane $\Pi_x$, which is a linear subspace of $T_x\mathbb{R}^3$, must satisfy:

$$0 = N(\mathbf{v}) = N_i v^i . \qquad (4.1)$$

One can replace the differential map $dx|_u$, which injects the tangent space $T_u S$ into $T_{x(u)}\mathbb{R}^3$, with the linear map $i_x : \mathbb{R}^2 \to T_x \mathbb{R}^3$, $(v^1, v^2) \mapsto \mathbf{v}$ that injects $\mathbb{R}^2$ into $T_x\mathbb{R}^3$ as the annihilating plane $\Pi_x$ for $N$ at $x$ in the tangent space $T_x\mathbb{R}^3$ to $x$. Relative to the canonical frame on $\mathbb{R}^2$ ([1]) and an *adapted frame* $\{\mathbf{e}_1, \mathbf{e}_2, \mathbf{e}_3\}$ for $\Pi_x$ (so $\mathbf{e}_1$ and $\mathbf{e}_3$ span $\Pi_x$, while $\mathbf{e}_3 \equiv \mathbf{N}$ is normal to it, and $\mathbf{v} = v^1 \mathbf{e}_1 + v^2 \mathbf{e}_2$), the matrix for the linear map $i_x$ will be $\delta^i_a$. When the frame on $\mathbb{R}^3$ is the natural frame, an adapted frame will take the form:

$$\mathbf{e}_i = x^j_i(x)\partial_j , \qquad \mathbf{e}_a = x^i_a(x)\partial_i \quad (a = 1, 2), \qquad \mathbf{e}_3 = x^i_3(x)\partial_i = \mathbf{N}, \qquad (4.2)$$

---

([1]) The canonical frame on $\mathbb{R}^3$ is defined by the three vectors $\delta_1 = (1, 0, 0)$, $\delta_2 = (0, 1, 0)$, $\delta_3 = (0, 0, 1)$.

in which $x_i^j(x)$ is an invertible 3×3 matrix whose components are differentiable functions of $x$. In the integrable case above, for which an embedding $x : S \to \mathbb{R}^3$ existed, we had:

$$x_a^i = \frac{\partial x^i}{\partial u^a}. \tag{4.3}$$

In what follows, we shall essentially replace all occurrences of the right-hand side of the equality (4.3) with the left-hand side, which no longer needs to take the form of a partial derivative matrix.

In order to make the normal vector $\mathbf{N}_x$ to each plane $\Pi_x$ more specific, since we have given $\mathbb{R}^3$ a metric, we define $\mathbf{N}$ to be the vector field that corresponds to $N$ under that metric:

$$\mathbf{N} = N^i \partial_i, \qquad N^i = \delta^{ij} N_j. \tag{4.4}$$

We then see that:

$$N(\mathbf{N}) = \delta(\mathbf{N}, \mathbf{N}) = 1, \tag{4.5}$$

i.e., $\mathbf{N}$ is a unit vector field.

We will then have:

$$d\mathbf{e}_i = dx_i^j \otimes \partial_j = \omega_i^j \otimes \mathbf{e}_j, \tag{4.6}$$

in which we have defined the matrix of 1-forms:

$$\omega_j^i \equiv dx_j^k \tilde{x}_k^i, \tag{4.7}$$

which agrees with the definition (3.11) for a surface, except for the nature of the matrix $x_j^i$.

That matrix can be expressed as $\partial_k x_j^l \tilde{x}_l^i dx^k$ with respect to the natural coframe field on $\mathbb{R}^3$ or in the form:

$$\omega_j^i = \omega_{jk}^i \theta^k, \qquad \omega_{jk}^i \equiv x_k^l \partial_l x_j^m \tilde{x}_m^i = (\mathbf{e}_k x_j^m) \tilde{x}_m^i \tag{4.8}$$

with respect to the adapted one. This definition of $\omega_{jk}^i$ also agrees with the one that was given for surfaces, except that now, in addition to a different definition of the matrix $x_j^i$, the vector field $\mathbf{e}_a$ does not have to act like $\partial / \partial u^a$.

We can then separate tangential components of $d\mathbf{e}_a$ from the normal one and expand the right-hand side of the last equality in (4.6) to give:

$$d\mathbf{e}_a = \omega_a^b \otimes \mathbf{e}_b + \omega_a^3 \otimes \mathbf{N}, \qquad d\mathbf{e}_3 = \omega_3^b \otimes \mathbf{e}_b + \omega_3^3 \otimes \mathbf{N}, \tag{4.9}$$

which shows that $d\mathbf{e}_a$ has both a tangential part $\omega_a^b \otimes \mathbf{e}_b$ and a normal part $\omega_a^3 \otimes \mathbf{N}$.

Note that the only things that have changed from the corresponding definitions on a surface is the replacement of the partial derivative matrix $x_{,a}^i$ with the more general one $x_a^i$ and the replacement of the normalized gradient of $f$ with the unit normal vector field $\mathbf{N}$.

The reciprocal coframe field $\theta^i$ to $\mathbf{e}_i$ [for which $\theta^i(\mathbf{e}_j) = \delta^i_j$] will then be represented by:

$$\theta^i = \tilde{x}^i_j(x)\,dx^j, \qquad \theta^a = \tilde{x}^a_j(x)\,dx^j, \qquad \theta^3 = \tilde{x}^3_j(x)\,dx^j \equiv N. \tag{4.10}$$

That coframe field will then obey the equations that are dual to (4.6):

$$d\theta^a = -\omega^a_b \otimes \theta^b - \omega^a_3 \otimes N, \quad d\theta^3 = -\omega^3_b \otimes \theta^b - \omega^3_3 \otimes N. \tag{4.11}$$

When the frame field $\mathbf{e}_i$ and its reciprocal coframe field are orthonormal, one can rewrite the equations for both fields as:

$$d\mathbf{e}_a = \omega^b_a \otimes \mathbf{e}_b + \omega^3_a \otimes \mathbf{N}, \qquad d\mathbf{N} = \omega^b_3 \otimes \mathbf{e}_b, \tag{4.12}$$

$$d\theta^a = -\omega^a_b \otimes \theta^b - \omega^a_3 \otimes N, \quad dN = -\omega^3_b \otimes \theta^b, \tag{4.13}$$

in which the matrix $\omega^i_j$ will be antisymmetric.

We can express $i^* d_\wedge N$ in terms of the adapted coframe $\theta^a$ in the forms:

$$i^* d_\wedge N = \tfrac{1}{2}(\partial_i N_j - \partial_j N_i)\,x^i_a x^j_b\,\theta^a \wedge \theta^b = \tfrac{1}{2}(\mathbf{e}_a N_i\,x^i_b - \mathbf{e}_b N_i\,x^i_a)\,\theta^a \wedge \theta^b \tag{4.14}$$

and

$$i^* d_\wedge N = -\tfrac{1}{2} N_i(\mathbf{e}_a x^i_b - \mathbf{e}_b x^i_a)\,\theta^a \wedge \theta^b = -\tfrac{1}{2} N[\mathbf{e}_a, \mathbf{e}_b]\,\theta^a \wedge \theta^b, \tag{4.15}$$

since:

$$[\mathbf{e}_a, \mathbf{e}_b] = (\mathbf{e}_a x^i_b - \mathbf{e}_b x^i_a)\partial_i. \tag{4.16}$$

Of course, since our pseudo-surface is not assumed to be integrable into a surface, we cannot assume that the normal component of $[\mathbf{e}_a, \mathbf{e}_b]$ vanishes anymore. Indeed:

$$[\mathbf{e}_a, \mathbf{e}_b] = c^c_{ab}\,\mathbf{e}_c + c^3_{ab}\,\mathbf{N}, \tag{4.17}$$

with

$$c^c_{ab} = (\mathbf{e}_a x^i_b - \mathbf{e}_b x^i_a)\,\tilde{x}^c_i, \qquad c^3_{ab} = (\mathbf{e}_a x^i_b - \mathbf{e}_b x^i_a)\,\tilde{x}^3_i. \tag{4.18}$$

With those preparations, we can then go about the business of defining the same things on pseudo-surfaces that were just defined for surfaces by simply substituting the injection $i_x$ for the differential map $dx|_u$, which will imply the substitution of $x^i_a$ for $\partial x^i / u^a$.

*b. Fundamental forms for pseudo-surfaces.* – The first fundamental form $g$ (i.e., metric) on the pseudo-surface $\Pi$ that is defined by the annihilating planes to a non-zero unit 1-form $N$ is the pull-back $i^*\delta$ of the Euclidian metric $\delta$ on $\mathbb{R}^3$ to $\Pi$ by way of the canonical injections $i_x$. If $\mathbf{v}, \mathbf{w} \in \Pi_x$ then:

$$g(\mathbf{v}, \mathbf{w}) = \delta(i_x(\mathbf{v}), i_x(\mathbf{w})). \tag{4.19}$$

Relative to the adapted coframe $\theta^i$:

$$g = g_{ab}\,\theta^a\theta^b + 2g_{a3}\,\theta^a\theta^3 + g_{33}\,\theta^3\theta^3, \quad \text{with} \quad g_{ij} = \delta_{kl}\,x^k_i\,x^l_j, \tag{4.20}$$

but $\theta^3(\mathbf{v}) = \theta^3(\mathbf{w}) = 0$, so $g$ will be simply:

$$g = g_{ab}\,\theta^a\theta^b, \quad \text{with} \quad g_{ab} = \delta_{ij}\,x^i_a\,x^j_b, \tag{4.21}$$

which should be compared with (3.23).

The second fundamental form of the pseudo-surface $\Pi$ is somewhat more involved, due to the fact that since $N$ is not exact, its differential:

$$dN = \partial_i N_j\,dx^i \otimes dx^j \tag{4.22}$$

will not have to be symmetric, in general.

However, one can polarize $dN$ into a sum of a symmetric part and an antisymmetric one using the permutation operator:

$$dN = d_s N + d_\wedge N, \tag{4.23}$$

in which we have defined:

$$d_s N = \tfrac{1}{2}(\partial_i N_j + \partial_j N_i)\,dx^i dx^j, \qquad d_\wedge N = \tfrac{1}{2}(\partial_i N_j - \partial_j N_i)\,dx^i \wedge dx^j. \tag{4.24}$$

Clearly, the antisymmetric part of the differential coincides with the exterior derivative, which would vanish if $N$ were exact.

Our approach to defining the second fundamental form of $\Pi$ is to pull back the symmetric part of $dN$ to $\Pi$ by once more using the canonical injections. Hence, if $\mathbf{v}, \mathbf{w}$ are tangent vectors in $\Pi_x$ then the second fundamental form $H_x = -\,i_x^*\,d_s N$ will be defined by:

$$H(\mathbf{v}, \mathbf{w}) = -\,d_s N(i_x(\mathbf{v}),\,i_x(\mathbf{w})). \tag{4.25}$$

Therefore, its components in an adapted coframe $\theta^a$ on $\Pi$ will be:

$$H_{ab} = -\tfrac{1}{2}(\partial_i N_j + \partial_j N_i)\,x^i_a\,x^j_b = -\tfrac{1}{2}(\mathbf{e}_a N_i\,x^i_b + \mathbf{e}_b N_i\,x^i_a), \tag{4.26}$$

which should be compared with the corresponding definition (3.28) for a surface. This time, $N_{i,j}$ is not generally symmetric and must therefore be symmetrized. In the final form, we have effectively replaced the partial derivative $\partial/\partial u^a = x^i_{,a}\,\partial_i$ (which does not exist now) with the directional derivative $\mathbf{e}_a = x^i_a\,\partial_i$ (which does).

We can adapt the conversion of $H_{ab}$ to an expression in terms of derivatives of $x_a^i$ in an analogous way to the one that was used above by starting from the fact that $N_i \, x_a^i = 0$ and taking the directional derivative this time:

$$0 = \mathbf{e}_a N_i \, x_b^i + N_i \, \mathbf{e}_b \, x_a^i, \qquad (4.27)$$

which will allow us to convert (4.26) into the form:

$$H_{ab} = N_i \, \mathbf{e}_{(a} \, x_{b)}^i, \qquad (4.28)$$

in which the parentheses indicate that the indices have been symmetrized; one can compare this expression with the previous one (3.30).

If we go back to the expression for $dN$ in (4.13) then we will once again have:

$$H_{ab} = \omega^3_{(ab)} \qquad (4.29)$$

for an orthonormal adapted coframe field.

The matrix $H_{ab}$ is still symmetric, though, and if the corresponding matrix $H^a_b = g^{ac} H_{cb}$ is also symmetric then it will again have real eigenvalues $\kappa_1$, $\kappa_2$, which can once more be identified with the principal curvatures of the pseudo-surface $\Pi$. The Gaussian and mean curvatures $K$ and $\bar{\kappa}$ are defined to be the determinant and one-half the trace of $H^a_b$, resp., as before, and curves in $\mathbb{R}^3$ whose tangent vectors lie along eigenvectors of $H^a_b$ will again be lines of curvature in the pseudo-surface.

*c. Curves on pseudo-surfaces.* – A curve $x(s)$ in $\mathbb{R}^3$ is said to *lie in the pseudo-surface* $\Pi$ if its tangent vectors $\mathbf{t}(s)$ always lie in the planes $\Pi_{x(s)}$. Hence, if the planes are annihilated by the 1-form $N$ then one will have:

$$N(\mathbf{t}(s)) = 0 \quad \text{for all } s. \qquad (4.30)$$

The components of $\mathbf{t}$ and $N$ with respect to the natural frame and coframe will then satisfy:

$$N_i \frac{dx^i}{ds} = 0. \qquad (4.31)$$

If we express the unit tangent vector field $\mathbf{t}(s)$ in the form $v^a(s) \, \mathbf{e}_a(s)$ then we will have:

$$\frac{d\mathbf{t}}{ds} = \frac{dv^a}{ds} \mathbf{e}_a + v^a \frac{d\mathbf{e}_a}{ds}. \qquad (4.32)$$

In order to get the derivative of $\mathbf{e}_a$ with respect to arc-length, we need only evaluate $d\mathbf{e}_a$, as it is expressed in (4.12), on $\mathbf{t}$:

$$\frac{d\mathbf{e}_a}{ds} = d\mathbf{e}_a(\mathbf{t}) = \omega_a^b(\mathbf{t})\mathbf{e}_b + \omega_a^3(\mathbf{t})\mathbf{N}, \tag{4.33}$$

which also has a tangent and a normal part. We can then substitute that in (4.32) and get:

$$\frac{d\mathbf{t}}{ds} = \left[\frac{dv^a}{ds} + \omega_b^a(\mathbf{t})v^b\right]\mathbf{e}_a + \omega_a^3(\mathbf{t})v^a\,\mathbf{N}. \tag{4.34}$$

The geodesic curvature of $x(s)$ is then:

$$\kappa_g = \left[\frac{dv^a}{ds} + \omega_b^a(\mathbf{t})v^b\right]\mathbf{e}_a, \tag{4.35}$$

while the normal curvature is:

$$\kappa_N = \omega_a^3(\mathbf{t})v^a = \omega_{(ab)}^3\, v^a v^a. \tag{4.36}$$

These results differ from the corresponding ones in (3.46) and (3.47) on surfaces only by the nature of the connection form $\omega_j^i$.

We again define a curve $x(s)$ in $\Pi$ to be a geodesic when its unit normal vector field $\mathbf{n}(s)$ is equal to $\mathbf{N}(x(s))$ for every $s$. From the Frenet-Serret equations:

$$\frac{d\mathbf{t}}{ds} = \kappa\mathbf{n}, \tag{4.37}$$

so if $x(s)$ is a geodesic then:

$$\frac{d\mathbf{t}}{ds} = \kappa\mathbf{N}, \quad \text{or} \quad \frac{d^2 x^i}{ds^2} = \kappa N^i. \tag{4.38}$$

That will be true iff:

$$\kappa_g = 0, \qquad \kappa_N = \kappa, \tag{4.39}$$

i.e.:

$$0 = \frac{dv^a}{ds} + \omega_b^a(\mathbf{t})v^b, \quad \kappa = \omega_{(ab)}^3\, v^a v^a, \tag{4.40}$$

which are essentially the same as for a surface, except for the difference in $\omega_j^i$.

In order to get $x(s)$ from this system, one first integrates the first set of equations for $v^a(s)$ and then integrates the first-order system:

$$\frac{dx^i}{ds} = x_a^i\, v^a \tag{4.41}$$

for $x^i(s)$.

One can differentiate the constraint (4.30) to get a corresponding constraint on $\mathbf{n}$:

$$0 = \frac{d}{ds}N(\mathbf{t}) = \frac{dN}{ds}(\mathbf{t}) + N\left(\frac{d\mathbf{t}}{ds}\right) = d_s N(\mathbf{t},\mathbf{t}) + \kappa N(\mathbf{n}),$$

in which we have defined $dN / ds$ by

$$\frac{dN}{ds} = i_t \, dN = i_t \, d_s N + i_t \, d_\wedge N , \qquad (4.42)$$

which would make:

$$\frac{dN}{ds}(\mathbf{t}) = i_t \, i_t \, d_s N + i_t \, i_t \, d_\wedge N = i_t \, i_t \, d_s N = d_s N \, (\mathbf{t}, \mathbf{t}) , \qquad (4.43)$$

since $i_t \, i_t \, d_\wedge N = d_\wedge N \, (\mathbf{t}, \mathbf{t})$ and $d_\wedge N$ is antisymmetric. Ultimately, the constraint on the normal vector field is:

$$\kappa N \, (\mathbf{n}) = - \, d_s N \, (\mathbf{t}, \mathbf{t}) . \qquad (4.44)$$

When $\mathbf{t}(s)$ is the injection of a curve $\upsilon(s) = \upsilon^a \, \partial_a$ in $\mathbb{R}^2$ by $i_x$, so:

$$t^i = x^i_a \, \upsilon^a , \qquad (4.45)$$

one will have:

$$\kappa N \, (\mathbf{n}) = H \, (\mathbf{t}, \mathbf{t}) = H_{ab} \, \upsilon^a \, \upsilon^b . \qquad (4.46)$$

[compare to (3.50)]

One also has that since the normal component of $d\mathbf{t} / ds$ is $\omega^3_{(ab)} \, \upsilon^a \, \upsilon^b$, one will once more we have:

$$\omega^3_{(ab)} \, \upsilon^a \, \upsilon^b = H_{ab} \, \upsilon^a \, \upsilon^b , \qquad (4.47)$$

which leads to:

$$\omega^3_{(bc)} = H_{bc} , \qquad (4.48)$$

and both of the last two equations are consistent with the corresponding ones (3.52) and (3.55) from the geometry of surfaces.

For a geodesic, (4.46) will become:

$$\kappa = H \, (\mathbf{t}, \mathbf{t}) , \qquad (4.49)$$

which is consistent with (3.57). Thus, the second fundamental form of a pseudo-surface has the same character as its analogue for surfaces.

*d. Motion on pseudo-surface*. – If one defines the time-parameterized curve $x(t)$ to lie in $\Pi$ by the condition that $\mathbf{v}$ must lie in $\Pi_{x(t)}$ for each $t$ then that will give:

$$0 = N \, (\mathbf{v} \, (t)) = v \, N \, (\mathbf{t} \, (s)) , \qquad (4.50)$$

so that is equivalent to saying the $\mathbf{t}(s)$ lies in $\Pi$, since $v$ must be non-zero if $s = s(t)$ is to be invertible.

If $\{\mathbf{e}_i, i = 1, 2, 3\}$ is adapted to $\Pi$, so $\{\mathbf{e}_a, a = 1, 2\}$ spans the constraint plane at each point along the curve $x(s)$ then the velocity vector must take the form:

$$\mathbf{v}(t) = \upsilon^a(t)\mathbf{e}_a \tag{4.51}$$

for some unique pair of $\upsilon^a(t)$, which will make the acceleration take the form:

$$\mathbf{a}(t) = \dot{\upsilon}^a \mathbf{e}_a + \upsilon^a \dot{\mathbf{e}}_a . \tag{4.52}$$

This time:

$$\dot{\mathbf{e}}_a = \frac{ds}{dt}\frac{d\mathbf{e}_a}{ds} = v\frac{d\mathbf{e}_a}{ds} = v[\omega_a^b(\mathbf{t})\mathbf{e}_b + \omega_a^3(\mathbf{t})\mathbf{e}_3],$$

which gives:

$$\dot{\mathbf{e}}_a = \omega_a^b(\mathbf{v})\mathbf{e}_b + \omega_a^3(\mathbf{v})\mathbf{e}_3 . \tag{4.53}$$

Thus:

$$\mathbf{a} = \left[\frac{d\upsilon^a}{dt} + \omega_b^a(\mathbf{v})\upsilon^b\right]\mathbf{e}_a + \omega_a^3(\mathbf{v})\upsilon^a \mathbf{e}_3 = \mathbf{a}_t + \mathbf{a}_N , \tag{4.54}$$

with

$$\mathbf{a}_t = \frac{d\upsilon^a}{dt} + \omega_b^a(\mathbf{v})\upsilon^b , \qquad \mathbf{a}_N = \omega_a^3(\mathbf{v})\mathbf{e}_3 = \omega_a^3(\mathbf{v})\upsilon^a \mathbf{N} . \tag{4.55}$$

Equation (4.54) is then then time-parameterized analogue of (4.34).

From the first equation in (4.55), the curve $x(t)$ will be a geodesic iff:

$$0 = \frac{d\upsilon^a}{dt} + \omega_b^a(\mathbf{v})\upsilon^b . \tag{4.56}$$

Now consider the normal component of the acceleration. When one applies the 1-form $N$ to $\mathbf{a}$, as in (3.63), one will get:

$$N(\mathbf{a}) = \dot{v}N(\mathbf{t}) + \kappa v^2 N(\mathbf{n}) = \kappa v^2 N(\mathbf{n}). \tag{4.57}$$

If one multiplies both sides of (4.44) by $v^2$ then one will now get the corresponding condition on acceleration:

$$N(\mathbf{a}) = H(\mathbf{v},\mathbf{v}) , \tag{4.58}$$

which agrees with (3.64).

When one applies $N$ to $\mathbf{a}$, as in (4.55), one will get:

$$N(\mathbf{a}) = \omega_a^3(\mathbf{v})\upsilon^a , \tag{4.59}$$

which implies that:

$$a^N = \kappa v^2 N(\mathbf{n}) = H(\mathbf{v},\mathbf{v}) = \omega_a^3(\mathbf{v})\upsilon^a , \tag{4.60}$$

so we have three equivalent ways of characterizing the normal acceleration: viz., the projection of the centripetal acceleration onto the normal $\mathbf{n}$ to the curve, which is $v^2$ times the curvature of the curve in the direction $\mathbf{v}$ [which is then $\kappa N(\mathbf{n})$], the expression $H(\mathbf{v},\mathbf{v})$, and the expression $\omega_a^3(\mathbf{v})\upsilon^a$, which comes from the equations of the moving frame $\mathbf{e}_a(t)$.

If we think of the pseudo-surface $\Pi$ as defining a non-holonomic constraint on motion in $\mathbb{R}^3$ then we can regard motion on $\Pi$ as non-holonomic motion in $\mathbb{R}^3$.

**5. The Foucault pendulum.** – In 1851, the French physicist (Jean Bernard) Léon Foucault (1819-1868) exhibited the rotation of the Earth by means of a freely-suspended pendulum in the Panthéon in Paris, although a similar device had been constructed by Vincenzo Viviani previously. The rotation of the Earth affected the pendulum by causing its plane of oscillation to precess at a rate that varied with the latitude of the pendulum.

*a. Basic theory.* – The basis for the precession of the plane of oscillation is the existence of the Coriolis acceleration that arises in a frame that is rotating with respect to an inertial frame.

Typically, an "inertial frame" is defined by fixing a spatial frame at some specific time point. For instance, the "M50 Earth-centered inertial frame" is defined on midnight of January 1, 1950 and has an *x*-axis along the Earth's line of equinoxes (the intersection of the equatorial plane and the plane of the ecliptic), which is positive towards the constellation Aries, a *z*-axis along the Earth's rotational axis that points towards the star Polaris, and a *y*-axis that is normal to the plane of the other two and oriented to make the frame right-handed. (That inertial frame is used for star almanacs, which list line-of-sight angles to various stars with respect to that frame.) The reason that the Earth-centered inertial frame has to be defined at a specific time point is that the rotational axis precesses around a cone whose semi-angle is 23.5º once every 26,000 years, as well as nutating with a smaller amplitude every 14 years; thus, the line of equinoxes also precesses and nutates in the plane of the ecliptic. Of course, over the time span of most observations and experiments, both of those corrections are typically negligible.

For the purposes of the Foucault pendulum, it is sufficient to consider a "laboratory" frame whose *xy*-plane is horizontal and whose *z*-axis is normal to that plane, and the unit vectors along those directions are oriented to define a right-handed orthonormal triad. One makes it "inertial" by fixing its position at some initial time and deals with the fact that as the Earth rotates, the laboratory frame will not only rotate, but also translate, by ignoring the translation.

In general, if the initial frame is inertial and takes the form of an orthonormal triad then one can integrate the vector fields (which are defined by parallel-translating the frame to all points) to give a rectangular coordinate system $\{x^1, x^2, x^3\}$ on the laboratory that makes the inertial frame take the form of the natural frame $\{\partial_1, \partial_2, \partial_3\}$ for the coordinate system. If another frame field $\{\mathbf{e}_1(t), \mathbf{e}_2(t), \mathbf{e}_3(t)\}$ is defined by applying the same time-varying rotation $R(t)$ to all frames of the inertial frame field then one can say that:

$$\mathbf{e}_i(t) = R_i^j(t)\partial_j. \tag{5.1}$$

Differentiation with respect to *t* will give:

$$\dot{\mathbf{e}}_i = \dot{R}_i^j \partial_j = \omega_i^j \mathbf{e}_j, \tag{5.2}$$

in which we have defined the time-varying matrix:

$$\omega_i^j = \dot{R}_i^k \tilde{R}_k^j, \tag{5.3}$$

which can be interpreted as the *angular velocity* of the moving frame with respect to the inertial one.

Note that there is nothing special about rotation matrices in this, except that they are invertible. That is, one could do the same thing with elements of the general linear group *GL* (*n*) in any dimension, as well. It just happens that the theory of rotational mechanics in three-dimensional space is more developed than the more general theory, although general relativity does generalize the picture somewhat. The main difference is that under polar decomposition, any invertible matrix will become the product of a rotation and finite strain. Furthermore, one can also use *GL*(*n*) to represent affine transformations, as well as Euclidian and Lorentzian rigid motions.

A second differentiation with respect to *t* will give:

$$\ddot{\mathbf{e}}_i = \omega_i^j \dot{\mathbf{e}}_j + \dot{\omega}_i^j \mathbf{e}_j = (\omega_k^j \omega_i^k + \dot{\omega}_i^j) \mathbf{e}_j. \tag{5.4}$$

Hence, in addition to the *angular acceleration* $\dot{\omega}_i^j$, one also has a term that is basically "centripetal" in character.

One might wonder where the Coriolis term comes from in the above calculations. Here, one encounters a subtlety of rotational mechanics: In order to get to the usual expressions for the relative velocity and acceleration of motion as viewed from a rotating frame, one must start with the so-called *radius (or position) vector field*:

$$\mathbf{r} = x^i \partial_i \tag{5.5}$$

that is defined by any coordinate system $(U, x^i)$. The purpose of this vector field is then to describe the positions of the points in space with respect to the chosen frame. In effect, one is "lifting" the coordinates of points in space to components of a vector field that vanishes at the origin of the coordinate system.

The reason that we say "so-called" is that **r** is not a true vector field unless the manifold on which the coordinate chart is defined is "flat" (or "affine"). That is because when one transforms to another coordinate system $y^i$, the coordinates $x^i$ usually transform by way of a *diffeomorphism* of the form $y^i = y^i(x^j)$, while the components $v^i$ of a vector field (being tangent objects) transform *linearly* by means of the differential map:

$$\bar{v}^i = \frac{\partial y^i}{\partial x^j} v^j. \tag{5.6}$$

The only time that a diffeomorphism is the same as its differential map is when the diffeomorphism is a linear map to begin with. However, the definition of a flat or affine manifold is that all of its coordinate transformations are linear (or more precisely, affine). One then sees why differential geometry makes little use of radius vector fields, but one cannot avoid them when one is reading the standard literature of rotational mechanics (and Fourier analysis, for that matter).

Now, differentiate **r** (*t*) in the form:

$$\mathbf{r}\,(t) = \bar{x}^i(t)\,\mathbf{e}_i\,(t)\,,\tag{5.7}$$

which will give, in succession:

$$\dot{\mathbf{r}} = (\omega^i_j\,\bar{x}^j + \dot{\bar{x}}^i)\,\mathbf{e}_i\,,\quad \ddot{\mathbf{r}} = [(\omega^i_k\omega^k_j + \dot{\omega}^i_j)\,\bar{x}^j + 2\omega^i_j\,\dot{\bar{x}}^j + \ddot{\bar{x}}^i]\,\mathbf{e}_i\,.\tag{5.8}$$

The components of the apparent ([1]) velocity and acceleration with respect to that frame are then:

$$\bar{v}^i = \omega^i_j\,\bar{x}^j + \dot{\bar{x}}^i \quad\text{and}\quad \bar{a}^i = (\omega^i_k\omega^k_j + \dot{\omega}^i_j)\,\bar{x}^j + 2\omega^i_j\,\dot{\bar{x}}^j + \ddot{\bar{x}}^i\,,\tag{5.9}$$

respectively.

One sees that the (relative) motion of the frame $\mathbf{e}_i\,(t)$ has contributed a term $\omega^i_j\,\bar{x}^j$ to the apparent velocity, while it has contributed the terms:

$$a^i_c = \omega^i_k\omega^k_j\,\bar{x}^j\,,\quad a^i_{\tan} = \dot{\omega}^i_j\,\bar{x}^j\,,\quad a^i_{\text{Cor}} = 2\omega^i_j\,\dot{\bar{x}}^j\tag{5.10}$$

to the apparent acceleration. The first one is basically the centripetal acceleration, the second is the tangential acceleration, and the last one is the *Coriolis acceleration*.

In the case of the Earth's rotation, the angular velocity $\omega_E$ is constant: namely, 1 rotation per day, which amounts to 72 microradians per second. Hence, the tangential acceleration will vanish. As for the centripetal acceleration, one sees that although it is proportional to $(\omega_E)^2$, which is quite small in magnitude, that basically gets multiplied by the distance from the point on the Earth's surface where the origin of the laboratory frame is defined to the rotational axis of the Earth, and that varies from 0 at the poles to 6378 km (6.378 ×10$^6$ m) at the equator. Hence, it has a maximum magnitude of 0.473×10$^{-2}$ m / s$^2$ or 2.52 milli-*g*'s. That is to be compared with 9.81 m / s$^2$ (1 *g*) of gravitational acceleration, which points essentially straight down, so it is normal to the surface.

The Coriolis acceleration is closely analogous to the Lorentz force on a charged point-particle moving in a magnetic field in that both are linearly dependent upon velocity. Indeed, in the previous article by the other [1], it was shown that the relativistically-invariant form of the Lorentz force could be represented as something that originated in the non-integrability of a constraint that is defined by a 1-form, namely, the electromagnetic potential 1-form *A* [as it is defined on a *U* (1) principal bundle]. That is, the charge is constrained to move in a four-dimensional pseudo-hypersurface, namely, the horizontal sub-bundle of the tangent bundle to the total space of the gauge bundle that is annihilated by the connection 1-form that *A* defines. The Lorentz force then comes from the Lie derivative of *A* along a horizontal curve.

The magnitude of the Coriolis acceleration is $2\omega_E\,v\,\sin\psi$, where $\psi$ is the latitude of the laboratory and *v* is the speed of the motion. Hence, the part of the product that is taken up by $2\omega_E \sin\psi$ will range from 0 at the equator to $2\omega_E$ at the poles. (It is 95.6 microradians per second where the author lives.) That means that when the maximum speed *v* of motion has order 1 m /s, the resulting acceleration will have a maximum magnitude of 9.76 micro-*g*'s where the author lives. Although that sounds negligible compared to gravity and even the centripetal acceleration, the difference is that the last two accelerations are mostly vertical in the laboratory frame, while the

---

 ([1])  The author has never liked the use of the adjective "fictitious" as a substitute for "frame-dependent," and has opted for the word "apparent" that was preferred by Lanczos [**9**]. There is certainly nothing fictitious (or even frame-dependent) about the disintegration of a grinding wheel whose angular velocity has greatly exceeded its safety limits.

Coriolis acceleration is horizontal. Furthermore, in one day (which is 86,400 seconds), the Coriolis acceleration would contribute 47.3° to the precession of the plane of oscillation of a pendulum at the author's latitude, which is certainly non-negligible.

Therefore, since we are looking at motion in a horizontal plane, we agree to ignore the centripetal contribution to the apparent acceleration, but not the Coriolis term. If $\bar{g}^i$ is the projection of the acceleration of gravitation that acts upon the pendulum ($-g \sin \alpha \, \mathbf{e}_1$, where the motion is along the $\mathbf{e}_1$ axis, and $\alpha$ is the angle between the suspension wire and the vertical) then what will be left is:

$$\ddot{\bar{x}}^i = 2\omega^i_j \, \dot{\bar{x}}^j + \bar{g}^i. \tag{5.11}$$

In Hamel's treatment of the topic [**6**], the form that these equations take for the projection onto the horizontal $xy$-plane of the motion of a pendulum that swings in a vertical plane that makes an angle of $\phi$ with the $xz$-plane is essentially:

$$\left. \begin{array}{l} \ddot{x} = -2(\omega_E \sin \psi) \, \dot{y} - g \sin \alpha \cos \phi, \\ \ddot{y} = \phantom{-}2(\omega_E \sin \psi) \, \dot{x} - g \sin \alpha \sin \phi. \end{array} \right\} \tag{5.12}$$

In this form, the components of velocity and acceleration are referred to the natural frame on the $xy$-plane.

We see that the matrix $\omega^i_j$ and the vector $\bar{g}^i$ will take the form:

$$\omega^i_j = \omega_E \sin \psi \begin{bmatrix} 0 & -1 \\ 1 & 0 \end{bmatrix}, \qquad \bar{g}^i = -g \sin \alpha \, (\cos \phi, \sin \phi), \tag{5.13}$$

resp., in that frame.

**6. The Foucault pseudo-surface.** – In order to reset the problem of motion in a rotating frame as motion with a non-holonomic constraint that is defined by a pseudo-surface, we first start by abstracting somewhat from the above situation. The basic constraint is that the motion of a pendulum will be described by motion along a line that is its projection onto the horizontal plane. Since conservation of angular momentum requires that the plane of the pendulum's motion must be constant in inertial space, one has that the aforementioned line of motion must rotate in the horizontal plane as a result of the Coriolis acceleration.

We illustrate our basic definitions in the following figure:

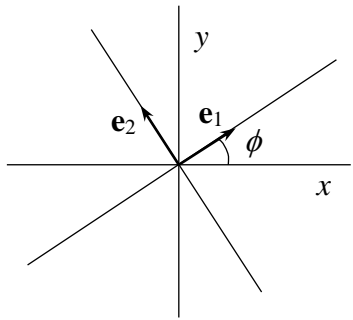

*a. The frame field defined by the Foucault pendulum.* – The laboratory frame will be represented by $\{\partial_x, \partial_y\}$. We define the moving frame $\{\mathbf{e}_1(t), \mathbf{e}_2(t)\}$ to be precessing in the *xy*-plane with a constant angular velocity of:

$$\dot{\phi} = 2\omega_E \sin \psi, \tag{6.1}$$

so

$$\phi(t) = \dot{\phi} t. \tag{6.2}$$

Since that makes:

$$\mathbf{e}_1 = \cos\phi \, \partial_x + \sin\phi \, \partial_y, \qquad \mathbf{e}_2 = -\sin\phi \, \partial_x + \cos\phi \, \partial_y, \tag{6.3}$$

our basic rotation matrices are:

$$R(t) = \begin{bmatrix} \cos\phi & -\sin\phi \\ \sin\phi & \cos\phi \end{bmatrix}, \qquad \tilde{R}(t) = \begin{bmatrix} \cos\phi & \sin\phi \\ -\sin\phi & \cos\phi \end{bmatrix}, \tag{6.4}$$

and the angular velocity 1-form is:

$$\omega = \tilde{R}\dot{R} = \dot{\phi}\begin{bmatrix} 0 & 1 \\ -1 & 0 \end{bmatrix} dt, \tag{6.5}$$

which differs from the one that was defined in (5.13) by the facts that the factor of 2 has been included in the definition of $\dot{\phi}$ and the sign has changed since the equations in (5.12) are expressed in terms of the frame $\{\partial_x, \partial_y\}$, which is rotating with the opposite sense with respect to the $\{\mathbf{e}_1, \mathbf{e}_2\}$.

The reciprocal coframe field $\{\theta^1, \theta^2\}$ to $\{\mathbf{e}_1, \mathbf{e}_2\}$ is defined by:

$$\theta^1 = \cos\phi \, dx + \sin\phi \, dy, \qquad \theta^2 = -\sin\phi \, dx + \cos\phi \, dy. \tag{6.6}$$

Since the rotation is time-varying, it becomes necessary to introduce time as another dimension to the configuration space. We set:

$$x^0 = t, \qquad \mathbf{e}_0 = \partial_t, \qquad \theta^0 = dt, \tag{6.7}$$

and extend the basic matrices to reflect the fact that the rotations do not affect the time axis:

$$R(t) = \begin{bmatrix} 1 & 0 & 0 \\ 0 & \cos\phi & -\sin\phi \\ 0 & \sin\phi & \cos\phi \end{bmatrix}, \quad \tilde{R}(t) = \begin{bmatrix} 1 & 0 & 0 \\ 0 & \cos\phi & \sin\phi \\ 0 & -\sin\phi & \cos\phi \end{bmatrix}, \quad \omega = \dot{\phi}\begin{bmatrix} 0 & 0 & 0 \\ 0 & 0 & 1 \\ 0 & -1 & 0 \end{bmatrix}\theta^0. \tag{6.8}$$

This means that the basic equations of the moving coframe can be written in the form:

$$\theta^i = \tilde{R}^i_j \, dx^j, \tag{6.9}$$

in which $i$ and $j$ range from 0 to 2. We can also infer the matrices $\omega^i_{jk}$ from the last equation in (6.8):

$$\omega^i_{j0} = \dot{\phi} \begin{bmatrix} 0 & 0 & 0 \\ 0 & 0 & 1 \\ 0 & -1 & 0 \end{bmatrix}, \qquad \omega^i_{jk} = 0, \qquad k = 1, 2. \tag{6.10}$$

Now that we have the components of $\omega^i_{jk}$, we can write out the differential equations of the adapted moving frame $\mathbf{e}_a(t)$ explicitly:

$$\frac{d\mathbf{e}_0}{dt} = \omega^b_{00} \mathbf{e}_b = 0, \qquad \frac{d\mathbf{e}_1}{dt} = \omega^b_{10} \mathbf{e}_b = -\dot{\phi} \mathbf{e}_2. \tag{6.11}$$

*b. Definition of the Foucault pseudo-surface.* – The constraint that we are imposing upon the motion is that it should always lie along the line generated by $\mathbf{e}_1$; i.e., in the space-time plane that is spanned by $\{\mathbf{e}_0, \mathbf{e}_1\}$. Since that plane is annihilated by the 1-form $\theta^2$, we have our field of tangent planes $\Pi$, which we shall refer to as the *Foucault pseudo-surface*. The frame field $\{\mathbf{e}_0, \mathbf{e}_1\}$ is then adapted to $\Pi$, as is its reciprocal coframe field $\{\theta^0, \theta^1\}$. Hence, the matrix of the injection of the constraint planes in the three-dimensional tangent spaces with respect to that frame is simply $\delta^i_a$ ($a = 1, 2$, $i = 0, 1, 2$). Furthermore, the adapted frame is also orthonormal since it was defined by a rotation of an orthonormal frame. The components $x^i_a$ of the adapted frame with respect to the natural frame field on $\mathbb{R}^3$ are then:

$$x^i_a = R^i_a = \begin{bmatrix} 1 & 0 \\ 0 & \cos\phi \\ 0 & \sin\phi \end{bmatrix}. \tag{6.12}$$

We see that the pseudo-surface that we have defined lives in the tangent bundle to a configuration manifold that is essentially $\mathbb{R}^3$ with the coordinates $(t, x, y)$ and consists of all the tangent planes that are annihilated by the 1-form $\theta^2$, which plays the role of $N$ above, so the integral submanifolds would be solutions to the Pfaff equation:

$$N = \theta^2 = 0. \tag{6.13}$$

In order to show that this Pfaff equation is not completely integrable, we first take the exterior derivative of $\theta^2$:

$$d_\wedge \theta^2 = d\tilde{R}^2_j \wedge dx^j = \dot{\tilde{R}}^2_j \, dt \wedge dx^j = -\omega^2_j \wedge \theta^j = -\omega^2_{10} \, \theta^0 \wedge \theta^1, \tag{6.14}$$

and with (6.10), that will give:

$$d_\wedge \theta^2 = \dot\phi \, \theta^0 \wedge \theta^1 . \tag{6.15}$$

We then compute the Frobenius 3-form that is defined by $\theta^2$:

$$\theta^2 \wedge d_\wedge \theta^2 = \dot\phi \, \theta^0 \wedge \theta^1 \wedge \theta^2 . \tag{6.16}$$

Since all three 1-forms involved are linearly independent, their exterior product (which amounts to a volume element on the configuration space) will not vanish, and the only way that the Frobenius form would vanish is if $\dot\phi$ itself vanished. Of course, the Earth would not be rotating then, so that would be out of the question. Thus, the non-integrability of the Foucault pseudo-surface originates in the rotation of the Earth.

Since the equation (6.13) is not completely integrable, there will be no constraint surfaces, but only constraint curves, i.e., the constraint is non-holonomic. It would be consistent to define the normal vector **N** to be $\mathbf{e}_2$, although the question of its orthogonality to Π requires some deeper analysis, which we shall get to later.

*c. Motion on the Foucault pseudo-surface.* – In order to discuss motion on the Foucault pseudo-surface, we must first extend the velocity of a curve in space to a curve in space-time.

Since $v^0 = dx^0 / dt = dt / dt = 1$, the most consistent way of extending the planar velocity vector $\mathbf{v} = v^1 \partial_x + v^2 \partial_y$ to a three-dimensional vector is:

$$\mathbf{v} = \partial_t + v^1 \partial_x + v^2 \partial_y = \mathbf{e}_0 + v^1 \mathbf{e}_1 + v^2 \mathbf{e}_2 . \tag{6.17}$$

When one includes the constraints that $v^0 = 1$ and $v^2 = 0$, and sets $v^1 = v$, the velocity of a curve that is constrained to the Foucault pseudo-surface will take the form:

$$\mathbf{v}(t) = \mathbf{e}_0 + v \, \mathbf{e}_1 \tag{6.18}$$

with respect to the adapted frame $\{\mathbf{e}_0, \mathbf{e}_1\}$.

We have actually introduced a second constraint by the demand that:

$$\theta^0(\mathbf{v}) = v^0 = 1 , \tag{6.19}$$

which implies that:

$$\dot v^0 = 0 . \tag{6.20}$$

However, since $\theta^0 = dt$, the constraint that we have introduced is completely integrable anyway, so we shall ignore it and simply focus on the one that is defined by $\theta^2(\mathbf{v}) = 0$. The way that we shall accomplish that it is to introduce the restrictions (6.19) and (6.20) in the final step of our calculations.

The acceleration of a curve that is constrained to lie in the Foucault pseudo-surface will then be:

$$\mathbf{a} = a^b \mathbf{e}_b + a^2 \mathbf{e}_2 = a^b \mathbf{e}_b + a^2 \mathbf{N} , \tag{6.21}$$

with:

$$a^b = \frac{dv^b}{dt} + \omega^b_{c0} v^c, \qquad a^2 = \omega^2_{c0} v^c, \qquad b, c = 0, 1 \; . \tag{6.22}$$

Explicitly, that gives:

$$a^0 = \dot{v}^0 = 0, \qquad a^1 = \dot{v} + \omega^1_{00} v^0 = \dot{v}, \qquad a^2 = \omega^2_{00} v^0 + \omega^2_{10} v^1 = -\dot{\phi} v, \tag{6.23}$$

when we introduce the restriction in (6.20).

From the equations of motion for the Foucault pendulum (5.12), the component $a^1$ of the acceleration in the direction of motion is:

$$a^1 = -g \sin \alpha = \dot{v}, \tag{6.24}$$

which is consistent with the fact that the speed decreases as the amplitude of the swing increases. The component $a^2$ of the acceleration that is normal to the line of motion in the spatial plane agrees with the Coriolis acceleration, since we have defined $\dot{\phi}$ to be $2\omega_E \sin \psi$, although the sign has changed with respect to the sign in the natural frame. Note that the curve described by the pendulum in our pseudo-surface would only be a geodesic if $\dot{v}$ were to vanish, which would happen in general only at the time points when $\alpha$ vanishes or if the pendulum were always vertical (i.e., $\alpha = 0$ for all time).

*d. The fundamental forms of the Foucault pseudo-surface.* – There is a subtle detail to address regarding the definition of the metric on $\mathbb{R}^3$: Up to now, we have been consistently assuming that our configuration manifold was $\mathbb{R}^3$ with the Euclidian metric. However, since we are not dealing $\mathbb{R}^3$ as model for "space" anymore, but a model for "space-time" (with the vertical dimension suppressed), we must be more careful about how we define a metric on it.

In non-relativistic physics, the metric on the tangent spaces is not actually the Euclidian one precisely or even a non-degenerate bilinear form. If one were to use the Euclidian metric to obtain the speed of the motion then one would get:

$$v^2 = 1 + (v^1)^2 + (v^2)^2, \tag{6.25}$$

in which the leading term would become somewhat superfluous.

A reasonable solution to the dilemma is to say that only the spatial planes spanned by $\{\partial_x, \partial_y\}$ or $\{\mathbf{e}_1, \mathbf{e}_2\}$ actually have metrics defined on them. That would make the metric on $\mathbb{R}^3$ take the form:

$$\bar{\delta} = dx^2 + dy^2. \tag{6.26}$$

In other words, the components that involve the time dimension would vanish, while the ones that pertain to spatial dimensions would coincide with the Euclidian metric:

$$\bar{\delta}_{0i} = 0, \qquad (i = 1, 2, 3), \qquad \bar{\delta}_{ij} = \delta_{ij} \qquad (i, j = 1, 2) \; . \tag{6.27}$$

$\mathbb{R}^3$, with the degenerate metric $\bar{\delta}$ that was just defined, is an example of what Yvonne Choquet-Bruhat [10] called a "Galilean embedding" of space in space-time.

When one uses the Galilean metric to convert the velocity vector into a covelocity covector, one will then get:

$$v = v_1\, dx^1 + v_2\, dx^2 \qquad (v_i \equiv \bar{\delta}_{ij}\, v^j), \tag{6.28}$$

but one will still have:

$$v(\mathbf{v}) = \bar{\delta}_{ij}\, v^i\, v^j = v^2. \tag{6.29}$$

Because the rotations performed involve only the spatial plane, they will still preserve the degenerate metric since the degeneracy is only in the time dimension. Hence, the expression for $\bar{\delta}$ in terms of the anholonomic coframe field $\theta^i$ on $\mathbb{R}^3$ will be:

$$\bar{\delta} = \bar{\delta}_{ij}\, \theta^i\, \theta^j = (\theta^1)^2 + (\theta^2)^2. \tag{6.30}$$

The matrix for the injection of the constraint planes (which are spanned by $\{\mathbf{e}_0, \mathbf{e}_1\}$) into the three-dimensional tangent spaces with respect to the adapted frame field will then be $\delta^i_a$ ($i = 1, 2, 3, a = 0, 1$), which will make the first fundamental form (viz., the metric on the pseudo-surface) have components:

$$g_{ab} = \bar{\delta}_{ij}\, \delta^i_a\, \delta^j_b = \begin{cases} 1 & a = b = 1, \\ 0 & \text{otherwise.} \end{cases} \tag{6.31}$$

Hence:

$$g = (\theta^1)^2. \tag{6.32}$$

In effect, since the motion is constrained to be spatially one-dimensional, we have extended that one-dimensional space to a two-dimensional space-time by a Galilean embedding.

An alternative to the Galilean embedding comes from relativistic physics, in which one uses a non-degenerate metric of Minkowski type:

$$\eta = \eta_{ij}\, dx^i\, dx^j, \quad dx^0 = c\, dt, \qquad \eta_{ij} = \text{diag}\, [+1, -1, -1]. \tag{6.33}$$

Since that metric relates to the proper time parameterization of a curve $x(\tau)$, the temporal component of velocity will then take the form:

$$v^0 = \frac{dt}{d\tau} = \left(1 - \frac{v^2}{c^2}\right)^{-1/2}, \tag{6.34}$$

in which $v$ is the norm of the spatial part of the velocity. Now, the problem of the Foucault pendulum does not usually involve relative velocities that are comparable to that of light, so $v^0$ will be essentially equal to 1, but at least one would have a non-degenerate matric to work with on the pseudo-surface, namely:

$$g = c^2 \, dt^2 - dy^2 = (\theta^0)^2 - (\theta^1)^2. \tag{6.35}$$

The second fundamental form is somewhat less ambiguous. One starts by computing the differential $dN = d\theta^2$ in the moving frame field on $\mathbb{R}^3$:

$$d\theta^2 = -\omega_j^2 \otimes \theta^j = -\omega_{10}^2 \, \theta^0 \otimes \theta^1 = \dot{\phi} \, \theta^0 \otimes \theta^1,$$

and symmetrizing it:

$$d_s\theta^2 = \dot{\phi} \, \theta^0 \, \theta^1. \tag{6.36}$$

[One could also get $d\theta^2$ from the second equation in (6.6) directly.] Since the tensor field $d_s\theta^2$ is already defined on the constraint planes, there is no need to project it, and that would make the second fundamental form equal to:

$$H = -\dot{\phi}\theta^0 \, \theta^1 = 2H_{01}\theta^0 \, \theta^1 \tag{6.37}$$

or

$$H = H_{ab}\theta^a \theta^b, \qquad H_{ab} = -\tfrac{1}{2}\dot{\phi}\begin{bmatrix} 0 & 1 \\ 1 & 0 \end{bmatrix} \qquad (a, b = 0, 1). \tag{6.38}$$

That makes the constraint on acceleration (4.58) take the form:

$$a^2 = -\dot{\phi}v^0 v^1 = -\dot{\phi}v, \tag{6.39}$$

which should be compared with the expression for $a^2$ in (6.23).

One can also check that $H_{ab} = \omega_{(ab)}^2$:

$$H_{00} = \omega_{00}^2 = 0, \qquad H_{01} = H_{10} = \tfrac{1}{2}(\omega_{01}^2 + \omega_{10}^2) = -\tfrac{1}{2}\dot{\phi}, \qquad H_{11} = \omega_{11}^2 = 0. \tag{6.40}$$

The problem comes when one attempts to find a principal frame for $H$. Namely, one cannot use the degenerate metric to raise an index in $H_{ab}$, because $g^{ab}$ does not exist. One can use the Minkowski metric to raise an index, for which $\eta^{ab} = \eta_{ab}$, but the resulting matrix will be:

$$H^a_{\ b} = H_{ab} = \tfrac{1}{2}\dot{\phi}\begin{bmatrix} 0 & -1 \\ 1 & 0 \end{bmatrix}, \tag{6.41}$$

which has only imaginary eigenvalues, namely, $\pm i\left(\tfrac{1}{2}\dot{\phi}\right)$, and therefore no real eigenvectors.

Of course, if one were to naively use the Euclidian metric on $\mathbb{R}^3$ then one would get $H^a_{\ b} = H_{ab}$, whose eigenvalues are real – namely, $\pm\tfrac{1}{2}\dot{\phi}$ – and one could put $H_{ab}$ into the form

$$H_{ab} = \tfrac{1}{2}\dot{\phi}\begin{bmatrix} 1 & 0 \\ 0 & -1 \end{bmatrix} \tag{6.42}$$

with respect to the principal frame, which would then describe an equilateral hyperboloid whose Gaussian curvature would be $-\tfrac{1}{4}\dot{\phi}^2$ and whose mean curvature would vanish; i.e., it would be a minimal surface. However, since the Euclidian metric on $\mathbb{R}^3$ has no physical significance, those consequences are somewhat moot.

*e. The Foucault pendulum and parallel translation.* – The precession of the plane of oscillation of a Foucault pendulum is sometimes explained (see, e.g., [**11**]) as being a result of the parallel translation of that plane (or really, its normal 1-form *N*) using a connection that is derived from the Coriolis acceleration. In order to see how parallel translation works in this case, the most direct route is to go back to the geometry of parallelizable manifolds – viz., teleparallelism.

When one has a global frame field $\mathbf{e}_i(x)$ on a parallelizable manifold ([1]), such as $\mathbb{R}^n$, if one *defines* the vector fields that comprise that frame field to each be parallel then one can say that any other vector field $\mathbf{v} = v^i \mathbf{e}_i$ is parallel iff its components $v^i$ with respect to that vector field are constants. Consequently, any other frame field:

$$\bar{\mathbf{e}}_i(x) = h_i^j(x)\mathbf{e}_j(x) \tag{6.43}$$

that is defined by a transition function $h_i^j(x)$ that is constant will define the same kind of parallelism for tangent vectors.

If we express the vector field **v** in terms of both $\mathbf{e}_i$ and the natural frame field $\partial_i$ on $\mathbb{R}^n$:

$$\mathbf{v} = v^i \mathbf{e}_i = \bar{v}^i \partial_i \tag{6.44}$$

and differentiate then we will get:

$$d\mathbf{v} = dv^i \otimes \mathbf{e}_i + v^i d\mathbf{e}_i = d\bar{v}^i \otimes \partial_i \tag{6.45}$$

or

$$d\mathbf{v} = (dv^i + \omega_j^i v^j) \otimes \mathbf{e}_i = d\bar{v}^j h_j^i \otimes \mathbf{e}_i . \tag{6.46}$$

That makes:

$$dv^i + \omega_j^i v^j = h_j^i d\bar{v}^j , \tag{6.47}$$

and since $v^i = h_j^i \bar{v}^j$, with a little rearranging we will get:

$$dv^i = h_j^i d\bar{v}^j + \omega_k^i h_j^k \bar{v}^j = h_j^i d\bar{v}^j + dh_j^i \bar{v}^j ,$$

---

([1]) There are topological obstructions to the definition of a global frame field on a differentiable manifold, but this is not the appropriate place to go into that. The author has, however, discussed those issues in another article on "singular teleparallelism" [**12**].

and when we take into account the facts that ($^1$):

$$\omega_j^i = dh_k^i \, \tilde{h}_j^k = -h_k^i \, d\tilde{h}_j^k = \tilde{h}_k^i \, dh_j^k = -d\tilde{h}_k^i \, h_j^k, \tag{6.48}$$

we will get:

$$dv^i = h_j^i (d\bar{v}^j + \omega_j^i \, \bar{v}^j). \tag{6.49}$$

Therefore, if the vector field **v** is parallel with respect to the frame field $\mathbf{e}_i$ (so $dv^i = 0$) then its components with respect to the natural frame field must satisfy:

$$0 = d\bar{v}^i + \omega_j^i \, \bar{v}^j, \tag{6.50}$$

which takes the form of the usual equations for the parallel translation of a vector field using the connection $\omega_j^i$.

Analogous considerations can be applied to the definition of a parallel covector field $\alpha = \alpha_i \, \theta^i$ when one defines the coframe field $\theta^i$ to also be parallel, which is consistent with the definition for $\mathbf{e}_i$. Namely, $\alpha$ is parallel iff $\alpha_i$ are all constants, and if $\alpha = \bar{\alpha}_i \, dx^i$ then $\alpha$ will be parallel iff:

$$0 = d\bar{\alpha}_i - \omega_i^j \, \bar{\alpha}_j. \tag{6.51}$$

The appearance of the minus sign on the right-hand side comes from the fact that the transformation from $dx^i$ to $\theta^i$ is "contragredient" to the transformation from $\partial_i$ to $\mathbf{e}_i$, i.e., it involves the inverse of the latter transformation.

Thus, any vector field or covector field that has constant components with respect to the precessing frame field that we have been using to describe the Foucault pendulum will have components in the inertial (i.e., natural) frame field that obey either (6.50) or (6.51), respectively. In particular, the motion of the pendulum, when projected onto the horizontal plane, will always lie along the line that is generated by $\mathbf{e}_1$, and if we define the teleparallelism connection to be the one that makes $\{\mathbf{e}_i, i = 0, 1, 2\}$ (*a fortiori*, $\{\mathbf{e}_a, a = 0, 1\}$) parallel then, by definition, $\mathbf{e}_1$ is being parallel-translated by the connection that comes from the Coriolis acceleration.

**7. Summary.** – Here are the basic results of this study:

1. We have defined the concept of a pseudo-surface in terms of the differential system on $\mathbb{R}^3$ that is defined by a Pfaff equation that is not completely integrable. Thus, although the differential system cannot be integrated into an actual surface, we can still define curves in a pseudo-surface, which take the form of motion in $\mathbb{R}^3$ that is subject to one non-holonomic linear constraint.

---

($^1$) The second-to-last equality comes about because when one replaces $h_j^i$ with its inverse, $\omega_j^i$, which is the infinitesimal generator of $h_j^i$, will go to its negative.

2. We have shown how the same geometric definitions (e.g., first and second fundamental form, Gaussian and mean curvature, geodesic equations) can be defined on a pseudo-surface that are defined on a surface and that the key to the analogy is in the replacement of an adapted coordinate system for a surface (and therefore an adapted holonomic frame field) with an adapted anholonomic frame field for a pseudo-surface.

3. We have shown how the Foucault pendulum defines such a pseudo-surface and exhibited its fundamental forms, as well as giving a physical interpretation of what geodesic motion would have to imply for a pendulum (i.e., no motion).

4. We have observed that trying to find principal curvatures of the Foucault pseudo-surface is straightforward for the Euclidian metric on $\mathbb{R}^3$, but meaningless for the Galilean metric, and that the Minkowski metric would give imaginary principal curvatures.

5. We have shown how the motion of the Foucault pendulum can be regarded as the parallel translation of its plane of oscillation by the teleparallelism connection that makes the rotating frame that is defined by the Coriolis acceleration parallel.

__________